\documentclass{ws-procs9x6}

\usepackage{feynmp}
\usepackage{subfigure}

\begin{document}

\title{BARYON RESONANCE PRODUCTION IN THE $\pi+d$ REACTION AND SEARCH FOR $\eta$-MESIC NUCLEI AT J-PARC }

\author{H. FUJIOKA}

\address{Department of Physics, Kyoto University,\\
Kyoto, 606-8502, Japan\\
E-mail: fujioka@scphys.kyoto-u.ac.jp}

\author{K. ITAHASHI}

\address{RIKEN Nishina Center for Accelerator-Based Science, RIKEN,\\
Saitama, 351-0198, Japan\\
E-mail: itahashi@riken.jp}

\begin{abstract}
A double-scattering reaction $\pi^+ + d\to p + p + \eta$, where an $\eta$ meson is rescattered, may provide new information on the $\eta N\to \eta N$ scattering. An idea to measure this reaction is proposed. Moreover, experimental search for $\eta$-mesic nuclei by the same $(\pi, \ N)$ reaction is discussed.
\end{abstract}


\bodymatter

\section{Introduction}

The possible existence of meson-nucleus bound systems is now one of the hot topics in hadron physics. Especially, an antikaon is expected to be bound in a finite nucleus, thanks to a strong attraction between an $I=0$ $\overline{K}N$ pair~\cite{Dote_HNP09}, and recent experimental findings of the signature of such a kaon-nucleus bound state~\cite{FINUDA05, DISTO09} have brought a heated debate on whether it really exists with a decay width narrow enough to be observed experimentally.

In contrast, the interaction between an $\eta$ meson and a nucleon are also known to be also attractive, while its strength is not well-determined. The $\eta N$ scattering length has a large ambiguity, between $0.270\,\mathrm{fm}$ and $1.050\,\mathrm{fm}$ for its real part and between $0.190\,\mathrm{fm}$ and $0.399\,\mathrm{fm}$ for its imaginary part~\cite{Haider02}. It is not possible to carry out neither an X-ray measurement of mesic atom nor a scattering experiment, in order to derive information on $\eta N$ interaction, different from the case for an antikaon. Instead, experimental data of hadronic production and photoproduction of $\eta$ mesons are mainly used in theoretical analyses. It should be noted that the $S_{11}(1535)$ resonance plays a dominant role in near-threshold $\eta$ production, since it lies just about $50\,\mathrm{MeV}$ above the $\eta N$ threshold, and strongly couples with the $\eta N$ channel, as seen in its large decay branching ratio into $\eta+N$ (45--60\%~\cite{PDG08}).

If the existence of $\eta$-mesic nuclei is confirmed experimentally, experimental observables such as the binding energy and decay width will impose a constraint on theoretical models about (in-medium) $\eta N$ interaction and the $S_{11}(1535)$ resonance. Haider and Liu predicted their existence for the first time~\cite{Haider86}, by using the scattering length of $(0.28+0.29i)\,\mathrm{fm}$ or $(0.27+0.22i)\,\mathrm{fm}$, and they found a bound state with the mass number $A \geq 12$ may exist. If the scattering length is larger than they applied, there may be a possibility for lighter $\eta$-mesic nuclei to exist.

In this paper, we discuss experimental ideas to explore $\eta N$ interaction and $\eta$-mesic nuclei by using the ($\pi$, $N$) reactions on different nuclear targets which will be feasible at the J-PARC facility.

\section{($\pi$, $N$) reaction at J-PARC}
Secondary particles, such as pions and kaons, are produced by the bombardment of intense proton beam accelerated by the J-PARC $50\,\mathrm{GeV}$ Proton Synchrotron onto the production target. Then, they are extracted and transported into experimental areas.

At present, there are two beamlines, K1.8 (available momentum up to $\sim 2\,\mathrm{GeV}/c$) and K1.8BR (up to $1.1\,\mathrm{GeV}/c$), and further beamlines are planned. For simplicity, we discuss experimental plans at the K1.8BR area, where most of the detectors for the E15 experiment~\cite{E15Prop} may be reused after a small modification.

\begin{figure}[t]
\begin{center}
\psfig{file=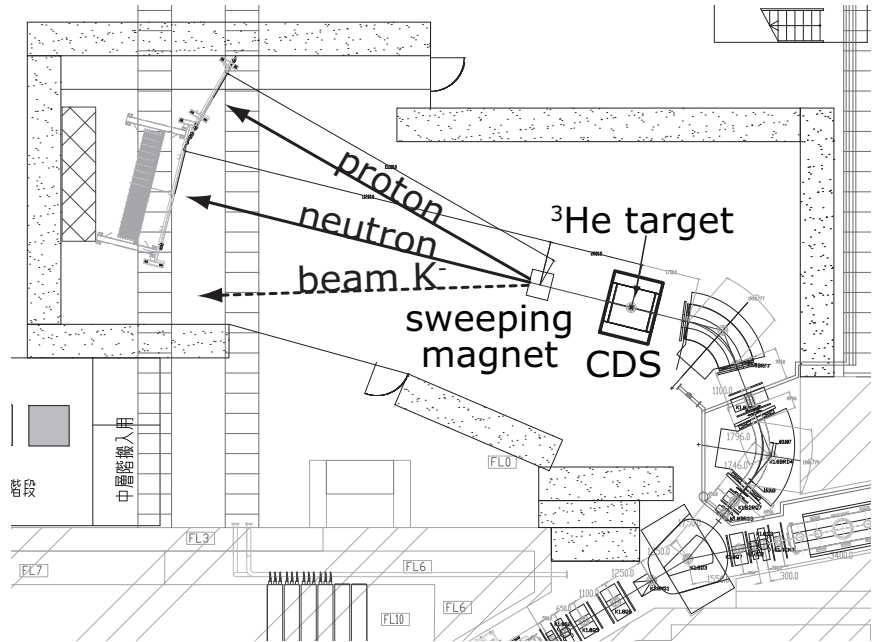,width=4.5in}
\caption{Setup for the J-PARC E15 experiment.}
\label{fig1}
\end{center}
\end{figure}

The E15 experiment aims to search for a deeply-bound kaonic nuclear state $K^-pp$ by the $^3\mathrm{He}$($K^-$, $n$) reaction. The layout of the experimental area together with the detectors to be installed is shown in Fig.~\ref{fig1}. A $K^-$ is transported through the K1.8BR beamline, and momentum-analyzed. Scattered neutrons by the ($K^-$, $n$) reaction on the helium-3 target are detected by the Neutron Counter, located $15\,\mathrm{m}$ downstream from the target. The decay particles of $K^-pp\to \Lambda +p\to p+\pi^-+p$ are detected by the Cylindrical Detector System which surrounds the target. Their trajectories can be reconstructed by the Cylindrical Drift Chamber installed in a solenoid magnet which operates at $0.5\,\mathrm{T}$ parallel to the beam axis. The Cylindrical Detector Hodoscopes, which locates outside of the CDC, are used for the trigger purpose and the particle identification.

Recently, a new measurement of the $^3\mathrm{He}$($K^-$, $p$) reaction in the E15 experiment, by adding a detector system for scattered protons, was proposed (J-PARC P28~\cite{P28Prop}), and the proposal has been approved as a part of the E15 experiment. Because a beam sweeping magnet is installed downstream of the target, scattered protons are also bent away from the beam axis (see Fig.~\ref{fig1}). Hence, an array of TOF counters for scattered protons will be located next to the charge veto counters for neutrons. In order to enable an inclusive measurement, a set of drift chambers will be installed downstream of the target, so that the trajectories of the protons can be reconstructed.

The present experimental setup is, of course, optimized for the E15 experiment. However, the ($\pi^\pm$, $n$) and/or ($\pi^\pm$, $p$) reactions on a different nuclear target can be investigated with almost the same setup. As discussed below, it is vitally important to detect rescattered particles in the $\pi^++d$ reaction and decay particles from $\eta$-mesic nuclei. Therefore, the usage of the Cylindrical Detector System will be desirable, too.

\section{Baryon resonance production in the $\pi+d$ reaction}
\begin{figure}[t]
\begin{center}
\subfigure[double-scattering reaction]{\psfig{file=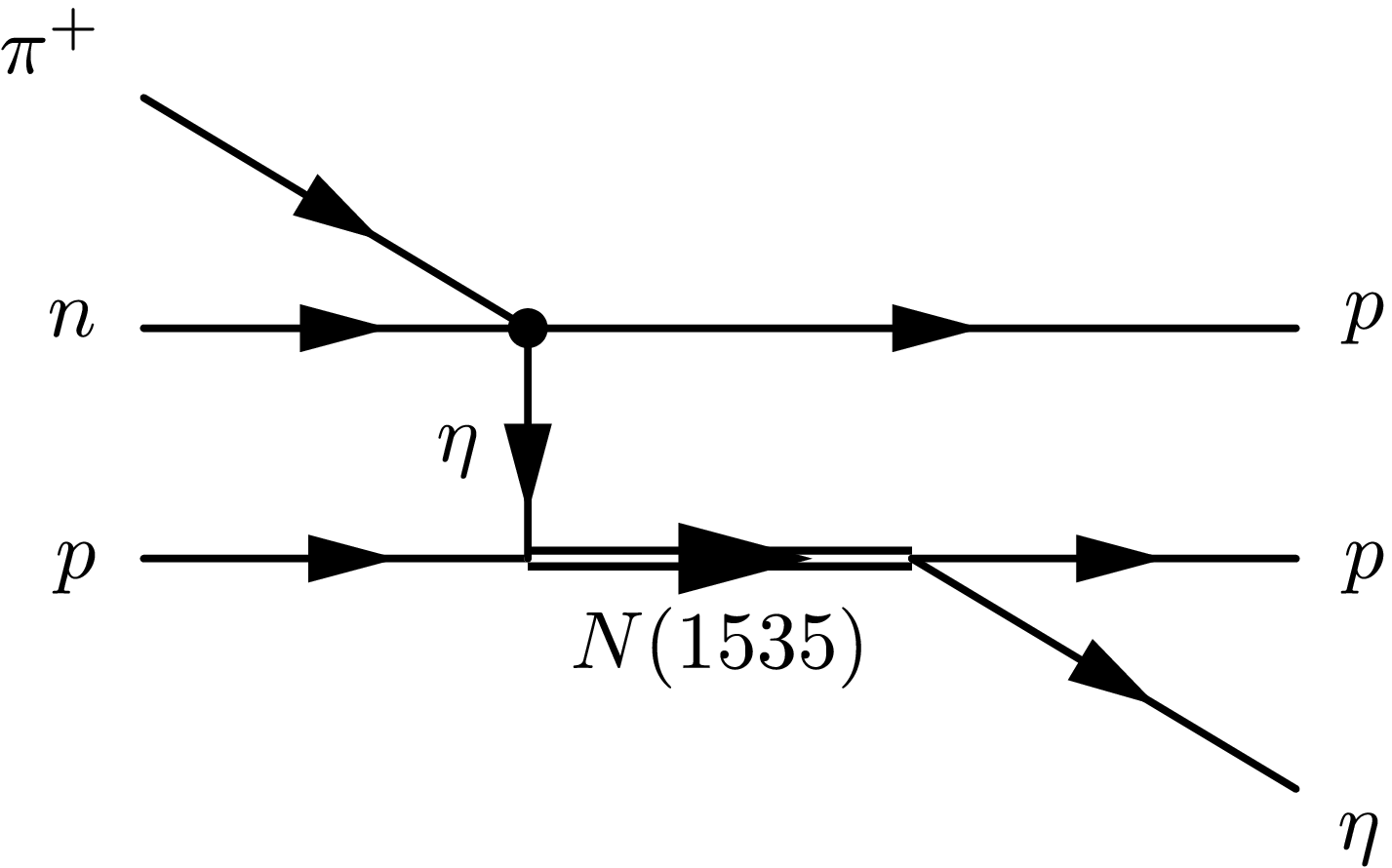,scale=0.35}\label{fig2a}}
\hspace{0.5cm}
\subfigure[quasi-free reaction]{\raisebox{9.9mm}{\psfig{file=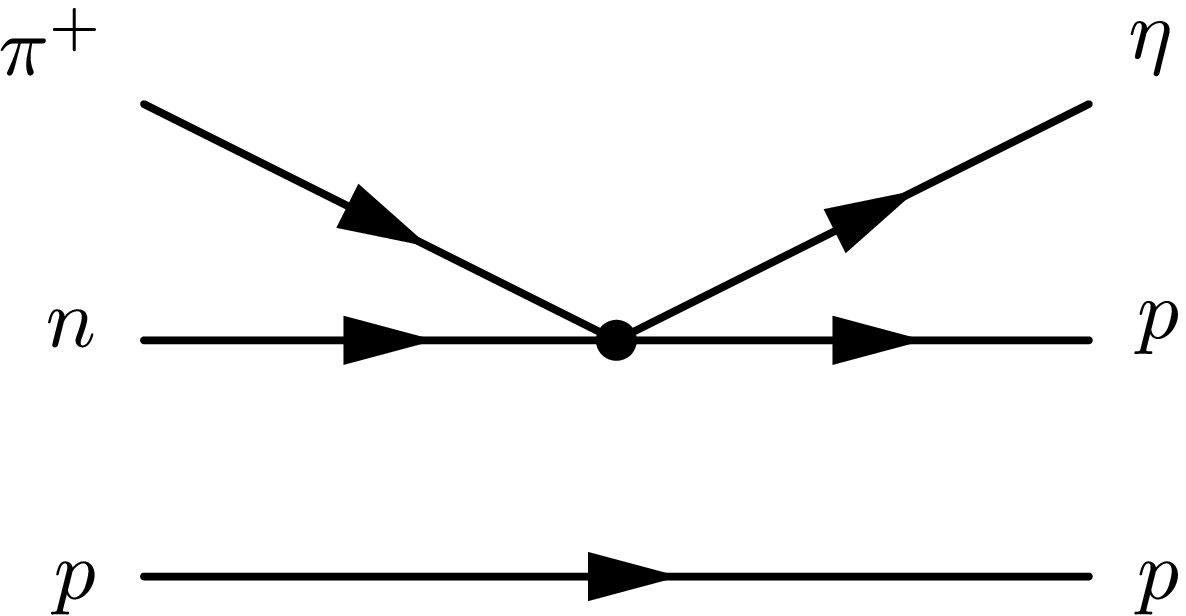,scale=0.35}}\label{fig2b}}
\caption{Diagrams for the $\pi^+d\to pp\eta$ reaction.}
\end{center}
\end{figure}

In this section, we consider a \textit{double-scattering reaction}:
\begin{equation}
\pi^++d \to p+p+\eta,\label{eq1}
\end{equation}
where both the protons in the final state are involved (Fig.~\ref{fig2a}). This reaction must be distinguished from a \textit{quasi-free reaction}:
\begin{equation}
\pi^+ +n \to p+\eta,\label{eq2}
\end{equation}
where the proton in the deuteron is a spectator. (Fig.~\ref{fig2b}). 

The double-scattering reaction is a two-step process, namely the quasi-free reaction (\ref{eq2}) followed by the rescattering:
\begin{equation}
\eta + p \to \eta + p.\label{eq3}
\end{equation}
The experimental goal is to obtain the missing-mass spectrum for the $d(\pi^+,\ p)X$ reaction with tagging $X$ as the $p\eta$ final state. It is expected that a bump corresponding to the $S_{11}(1535)$ resonance due to the rescattering (\ref{eq3}) will be observed in the spectrum.

\begin{figure}[t]
\begin{center}
\psfig{file=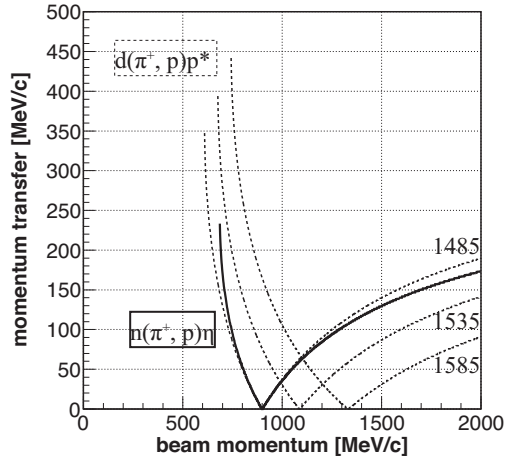,width=2.6in}
\caption{Momentum transfer for the $n$($\pi^+$, $p$)$\eta$ reaction (solid line) and the $d$($\pi^+$, $p$)$p^*$ reaction (dashed lines) for three different $p^*$ masses (1485, 1535, 1585$\,\mathrm{MeV}/c^2$).}
\label{fig3}
\end{center}
\end{figure}

Since we are interested in $\eta N$ interaction at low energy, the beam momentum should not be far from the recoilless condition for the $n(\pi^+,\ p)\eta$ reaction. As shown in Fig.~\ref{fig3}, the optimal beam momentum will be around $0.8\, \mbox{--}\,1.1\,\mathrm{GeV}/c$.
The momentum of the $S_{11}(1535)$ ($p^*$) resonance will be almost at rest (Fig.~\ref{fig3}), as expected.

Then, one possibility to separate the double-scattering reaction from the quasi-free one is to detect a proton from the $S_{11}$ decay ($p^*\to p+\eta$). While a proton from the decay of a $S_{11}$ with its mass $1.535\,\mathrm{GeV}/c^2$ has its momentum of $\sim 186\,\mathrm{MeV}/c$, the Fermi momentum of a nucleon in the deuteron hardly exceed $200\,\mathrm{MeV}/c$. Moreover by detecting a decay proton, a missing $\eta$ can be identified by use of the missing-mass of the $d(\pi^+,\ pp)X$ reaction.

From an experimental point of view, the present CDC for the E15 experiment does not allow for the detection of such a slow proton because of the existence of its inner support. For instance, the range of a $200\,\mathrm{MeV}/c$ proton is only $5\,\mathrm{mm}$ in a plastic scintillator. Accordingly, a target has to be thin enough, less than a few mm\footnote{Therefore, a solid $\mathrm{CD}_2$ will be used, instead of liquid $\mathrm{D}_2$.}, and we have to equip a new detector system between the CDC and the target. For example, it is considered to install two barrels of plastic scintillators with different distances from the center; the counter in the inner barrel measures the total energy of a proton, and that in the outer barrel serves as a veto counter to confirm the proton is stopped in the inner counter.

By the way, the $\pi^-+p\to \eta + n$ reaction has been investigated by several groups~\cite{pipetan, Prakhov}. The total cross section increases very rapidly from the threshold (center-of-mass energy $W\sim 1.487\,\mathrm{GeV}$), and reaches more than $2\,\mathrm{mb}$ in the $S_{11}(1535)$ resonance region. While the bump below $W\sim 1.6\,\mathrm\,\mathrm{MeV}$ can be explained by the predominant contribution of the $S_{11}(1535)$ resonance, the cross section above $1.6\,\mathrm{GeV}$ is strongly model-dependent~\cite{pipetan_th}. It is because more resonances, such as $S_{11}(1650)$, $P_{11}(1710)$, $D_{13}(1520)$, $D_{13}(1700)$, $P_{13}(1720)$, and so on, contribute to the reaction. 

A similar behavior should hold for the $\pi^++n\to \eta + p$ reaction, due to the isospin symmetry. Therefore, an incident pion momentum around $0.8\,\mathrm{GeV}/c$, which corresponds to $W\sim 1.56\,\mathrm{GeV}$, may be suitable in order to avoid difficulties in model calculations. A measurement with a higher beam momentum may provide information on different resonances, including the second $S_{11}$ resonance.

It should be noted that not only $\eta$ but also $\pi$ and vector mesons ($\rho$, $\omega$) can propagate as an intermediate meson, even if the center-of-mass energy of the pion and the neutron is adjusted to the $S_{11}(1535)$ region.  This is also the case for the $p+p\to p+p+\eta$ reaction near the threshold, which has been experimentally studied by COSY-11 and COSY-TOF (see, for example, Ref.~\refcite{Moskal09}). Theoretical calculations for the $p+p$ reaction~\cite{pp_th} take into account these intermediate mesons between two nucleons.

Therefore, a new calculation for the $\pi^++d$ reaction, based on a model for the $\pi^++p$ and $p+p$ reactions, is awaited. This novel reaction might reveal unique features, i.e., a possibility to make a constraint to parameters such as the a $MNN^*$ coupling constant.

It is worth noting that an analogous reaction has been already discussed with regard to the $\Lambda(1405)$ resonance. In the chiral unitary model, it is shown that a $\Lambda(1405)$, which is dynamically generated in the meson-baryon scattering, is a superposition of two poles~\cite{Jido03}. The lower pole ($\sim 1390\,\mathrm{MeV}$) and higher pole ($\sim 1420\,\mathrm{MeV}$) couple strongly with the $\pi\Sigma$ and $\overline{K}N$ channel, respectively. In order to confirm that the resonance in the $\overline{K}N$ channel locates around $1420\,\mathrm{MeV}$ instead of $1405\,\mathrm{MeV}$, the scattering of $\overline{K}N\to\pi\Sigma$ has to be examined. However, it is impossible to produce the $\Lambda(1405)$ resonance in a direct $K^-+p$ scattering, because it lies below the $\overline{K}N$ threshold. Jido~\textit{et al.}~\cite{Jido09} discussed a two-step process of the $K^-+d \to \Lambda(1405) +n$ reaction, where the resonance can be produced by the scattering of an nucleon in the deuteron and an intermediate antikaon, and compared their calculation with an old experiment by Braun \textit{et al.}~\cite{Braun77}\footnote{Since they used a deuteron bubble chamber, the scattered neutron was not detected. Instead, they obtained the $\pi^+\Sigma^-$ invariant-mass spectrum.}\footnote{Not only $\Lambda(1405)+n$, but also $\Sigma^-+p$, $\Sigma^-(1385)+p$, and $\Lambda(1520)+n$ were identified as final states.} In J-PARC, an experimental proposal to investigate the $d(K^-,\ n)$ reaction, together with the identification of the decay mode of $\Lambda(1405)$ into $\pi^0\Sigma^0$ as well as $\pi^\pm \Sigma^\mp$ has been submitted~\cite{P31Prop}.

Turning back to the double-scattering reaction (\ref{eq1}), the situation is different in that the $S_{11}(1535)$ resonance is above the $\eta N$ threshold. Nevertheless, there is a possibility that one can access information on the $\eta N\to \eta N$ scattering, caused by an intermediate $\eta$ meson, which cannot be investigated by a direct reaction with ``$\eta$ meson beam''.

\section{Search for $\eta$-mesic nuclei}
After Haider and Liu predicted the existence of $\eta$-mesic nuclei in 1986~\cite{Haider86}, $\eta$-nucleon and $\eta$-nucleus interaction have been of particular interest both theoretically and experimentally. Observation of $\eta$-mesic nuclei, as well as investigation of near-threshold $\eta$ production, will provide much better understanding of the interaction.

The TAPS/MAMI experiment investigated the photoproduction of $\eta$-mesic $^3\mathrm{He}$ by the $\gamma +{}^3\mathrm{He}\to \eta+X$ and $\pi^0+p+X$ reaction~\cite{Pfeiffer04}. The total cross section of the coherent $\eta$ production exhibits a peak-like structure at the threshold. An enhancement of correlated $\pi^0 p$ pairs with relative angles near $180^\circ$ is also observed near the $\eta$ production threshold, which may be attributed to the decay of $\eta$-mesic nuclei. By a simultaneous fit of the cross sections for the two channels, the binding energy $-4.4\pm 4.2\,\mathrm{MeV}$ and the width $25.6\pm 6.1\,\mathrm{MeV}$ are obtained.
However, Hanhart claimed that it is not possible to conclude whether the observed enhancement corresponds to a bound state or a virtual state, due to the lack of statistics~\cite{Hanhart05}.

Recently, the COSY-GEM experiment investigated the ${}^{27}\mathrm{Al}(p,\ {}^3\mathrm{He})$ reaction at the recoilless kinematics~\cite{COSYGEM09}. They observed back-to-back $\pi^-$-$p$ pairs, whose energy corresponds to $S_{11}(1535)$ at rest. The missing-mass spectrum, with requiring back-to-back $\pi^-$-$p$ pairs in coincidence, shows an enhancement below the $^{25}\mathrm{Mg}+\eta$ threshold. They obtained the binding energy ($-13.13\pm  1.64\,\mathrm{MeV})$ and the width ($10.22\pm 2.98\,\mathrm{MeV}$) of the ${}^{25}\mathrm{Mg}\otimes \eta$ system.

Both the experiments tagged a pair of a pion and a nucleon as the decay product of $\eta$-mesic nuclei. Since an $\eta$ meson and a nucleon strongly couples with a $S_{11}(1535)$ resonance ($N^*$), a main decay channel is considered to be $N^*\to N\pi$. The $N^*\to N\eta$ decay, which has a large branching ratio in free space, is suppressed because of the Pauli blocking, even if the state is above the $\eta$ emission threshold. Other decay modes, namely $N^*N\to NN$ and $NN\pi$, also contributes to the total decay width. The width of $NN$ decay is estimated to be negligibly small~\cite{Chiang91, Nagahiro03}. Experimentally, the $N^*\to N\pi$ is easiest to identify, as its back-to-back topology is a clean signal.

Historically, the first experiment to search for $\eta$-mesic nuclei was done at BNL by using the $(\pi^+,\ p)$ reaction on lithium, carbon, oxygen, and aluminum targets~\cite{Chrien88}, motivated by a theoretical calculation by Liu and Haider~\cite{Liu86}. A peak structure corresponding to $\eta$-mesic nuclei was not observed for any target.

We consider that it is worthwhile to reconsider this reaction for two reasons, and plan to carry out a new experiment at J-PARC~\cite{Itahashi_LoI}.

\begin{figure}[t]
\begin{center}
\psfig{file=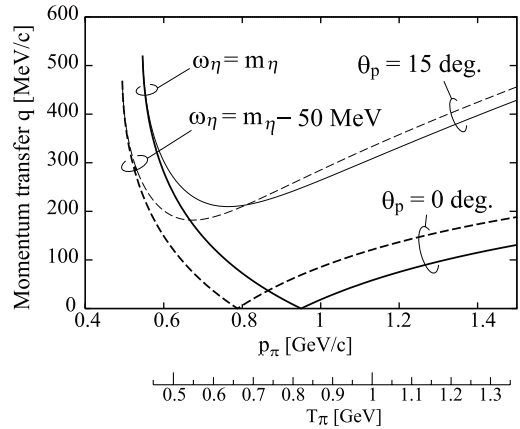,width=2.75in}
\caption{Momentum transfer for the $A(\pi,\ N)$ reaction (heavy mass limit) as a function of incident pion momentum ($p_\pi$) or kinetic energy ($T_\pi$), for different scattering angle ($\theta_p$) of $0^\circ$ and $15^\circ$. The solid (dashed) line corresponds to the $\eta$ energy at $\omega_\eta=m_\eta$ ($m_\eta-50\,\mathrm{MeV}$). This figure is taken from Ref.~\refcite{Nagahiro09}.}
\label{fig4}
\end{center}
\end{figure}

First, the proton spectrometer was set to $15^\circ$ against the beam in the BNL experiment, and the momentum transfer is as large as $200\,\mathrm{MeV}/c$ for the incident beam momentum $800\,\mathrm{MeV}/c$ (see Fig.~\ref{fig4}). A theoretical calculation by Nagahiro \textit{et al.} revealed that many subcomponents of different configurations of a nucleon-hole and an $\eta$ meson will contribute with substantial fractions, and that they will mask a possible peak structure of $\eta$-mesic nuclei~\cite{Nagahiro09}. 

On the other hand, the recoilless condition can be satisfied by measuring a proton scattered at almost zero degree. As shown in Fig.~\ref{fig4}, the magic momentum, for the $\eta$ energy $\omega_\eta=m_\eta-50\,\mathrm{MeV}$ and $m_\eta$, where $m_\eta$ is the mass of an $\eta$ meson, is $777\,\mathrm{MeV}/c$ and $950\,\mathrm{MeV}/c$, respectively.  In this condition only substitutional states are enhanced, since the angular momentum of the target nucleus is conserved after the reaction.

Secondly, the BNL experiment investigated only the inclusive reaction. A signal of $\eta$-mesic nuclei, if exists, may be overwhelmed by a large background irrelevant to $\eta$ production. Such a background will be reduced by detecting $N\pi$ pairs in back-to-back simultaneously.

Taking these points into consideration, we proposed to study the $(\pi^-,\ n)$ reactions on $^7\mathrm{Li}$ target at J-PARC~\cite{Itahashi_LoI} (see also Ref.~\refcite{Nagahiro09}). We have chosen $^7\mathrm{Li}$, in addition to $^{12}\mathrm{C}$ as a candidate of the target, because of less contribution of a proton in the $p_{3/2}$-shell. It is replaced by an $\eta$ meson with its orbital angular momentum 1 after the reaction ($p$-wave component), and it is foreseen that the structure from the $s$-wave component from a proton in the $s_{1/2}$-shell is much more interesting than that from the $p$-wave component.

Furthermore, the missing-mass spectrum will be affected by the in-medium property of the $N^*(1535)$ resonance. Nagahiro \textit{et al.} derived the $\eta$-nucleus optical potentials, based on two kinds of chiral models --- \textit{chiral doublet model} and \textit{chiral unitary model} --- and calculated the spectrum for each potential~\cite{Nagahiro09}.

In the chiral doublet model, the $N^*$ is regarded as the chiral partner of the nucleon, and the mass gap between $N^*$ and $N$ decreases in a finite density because of partial restoration of chiral symmetry. Thereby, a level crossing between the $N^*$-$N^{-1}$ and $\eta$ modes may take place in a nucleus, which results in a deeply-bound $\eta$ state and a bump structure in the unbound region.

On the other hand, the mass gap will not change largely in nuclear medium in case of the chiral unitary model. A shallow bound state, which is similar to the result in Ref.~\refcite{Garcia02}, will appear in the missing-mass spectrum.

Therefore, we expect to obtain information on the in-medium $N^*$ property in relation with the level crossing, via a missing-mass spectroscopy covering the unbound region as well as the bound region.

\section{Conclusion}
We discussed experimental ideas to utilize the $(\pi,\ N)$ reactions for two purposes: investigation of the properties of the $S_{11}(1535)$ resonance, and search for $\eta$-mesic nuclei. It seems that both studies will shed light on the $\eta$-nucleon and $\eta$-nucleus interaction.

\section*{Acknowledgments}
We acknowledge stimulating comments from Dr. M. D\"{o}ring. We would like to thank Dr. H.~Nagahiro, Dr. D.~Jido, and Prof. S.~Hirenzaki for valuable discussions on $\eta$-mesic nuclei.

\end{document}